\documentclass[journal,onecolumn]{IEEEtran}
\ifCLASSINFOpdf
   \usepackage[pdftex]{graphicx}
  \graphicspath{{../pdf/}{../jpeg/}}
\else
\fi
\begin{document}
\title{Analysis of Ensemble Forecasting Technique for Photovoltaic Power Generation }

\author{Rustam Kumar

Department of Electrical Engineering

Indian Institute of Technology Kanpur

Kanpur-208016, India

Email: rustamkumar22@gmail.com% <-this % stops a space
\thanks{This work is part of course EE632A ( Instructor:-Dr. S.C Srivastava, scs@iitk.ac.in)}% <-this % stops a space
}

\maketitle

\begin{abstract}
To cater the rapidly growing demand for electricity leading to the integration of renewable energy sources in power system. Due to intermittent nature of renewables, it also brings challenges for research community during the planning and operation stage in power system. Therefore it is primary necessity of the community to develop an accurate forecasting technique to solve the intermittency problem. In this report, A forecasting technique is proposed based on ensemble of state of the art forecasting techniques. For performance comparison among the techniques, GEFCom2014 meteorological data are used to predict the photovoltaic power, and the obtained results are included in this report.  
\end{abstract}

%\begin{IEEEkeywords}
%IEEE, IEEEtran, journal, \LaTeX, paper, template.
%\end{IEEEkeywords}

% \fi
\IEEEpeerreviewmaketitle

\section{Introduction}
\IEEEPARstart{H}{igh} penetration level of renewable generation in the power system introduces several challenges due to intermittent and uncertain nature [1]-[5]; therefore researchers are trying to predict the nature of renewable generation using forecasting techniques.  Over a few years, many techniques have been developed based on physical, statistical and ensemble methods. 
 The physical methods are a direct way of forecasting the output power. However, it requires understanding of device physics to derive mathematical models. These models are simple if global irradiance are only considered as input variable. However, accuracy of model is compromised. To obtain more accurate model, additional input variables are also included but the model becomes more complex [6]. Whereas the statistical methods are based on finding the relation between input variables (meteorological data) and output parameter (power) [7]. Therefore it doesn’t require any information related to Photovoltaic (PV) system such as location, PV model, etc.. However, it requires historical dataset during training. 
 
The state of the art forecasting methods is Artificial Neural Network (ANN), K- Nearest Neighbours (kNN), Support Vector Regression (SVR), Quintile Random Forest (QRF) and Ensemble Averaging (ENS) [8].  The performance of these methods depends on weather conditions like, during sunny days the kNN performs well, where QRF performs well during the intermediate value of Clear Sky Index (CSI). The detailed analysis is discussed in literature [8]. In the literature [8], an ENS method is also discussed, which uses Grey Box (GB), Neural Network (NN), kNN, QRF, and SVR methods, and it observed that the method performs well in all the provided weather condition. However, The ENS method is computationally intensive because using all the above methods. 

In this report, an ENS method is proposed based on the ensemble of kNN, SVR and QRF. The GB and NN methods are not considered because of poor performance, which makes this ENS method to be less computational extensive. The performance of proposed method is evaluated and compared with state of the art. The report is organised as follows: In section II the proposed methods are discussed, Results are presented in Section III, finally conclusion is drawn in section IV.

\section{Methodology}
\subsection{Neural Network (NN)}
During the developing of NN, we have used a feed-forward NN method. The architecture of the network is selected as per literature suggestion [8]. In the literature, after trying possible combinations between the hidden layer, activation function and regularization methods found the performance of network is optimal for a particular architecture. Therefore, we have also selected the network with a hidden layer of having 3 neurons with sigmoidal activation function. For training of the network, we have also considered Bayesian regularization method. For simplicity we have kept the architecture static during the test. Wherever we have weekly updated the weight for good network performance.
\subsection{K- Nearest Neighbours (kNN)}
The kNN method is simplest methods in the ML, this methodology uses the idea of similarity like; For given weather condition, it searches for k similar weather condition from the previously available datasets and then the corresponding output power is combined using weighted average method keeping the most weight to closest historical data. The challenges of the network lie in choosing parameters like normalizing factors for the dataset, computing the distances between predicted and historical data, considering the no of neighbors and giving weight to their corresponding output power. In this work, we have also used the same parameters as discussed in literature [8]. All the meteorological variable is normalized between [0-1] and the Euclidean distance is considered for distance calculation. A total 300 neighborhoods are considered and weighted average is applied according to Gaussian similarity kernel.
In this case, we have used 5 meteorological variables. That is strongly coupled to the power such as Total Cloud Cover (TCC), Surface Solar Radiation Down (SSRD), Surface Thermal Radiation Down (STRD), Top net Solar Radiation (TSR) and Total Precipitation (TP). It is found that using the more no of input variables than NN network, this network performs well than NN network.

\begin{figure}[h!]
\centerline{\includegraphics[width=0.6\textwidth]{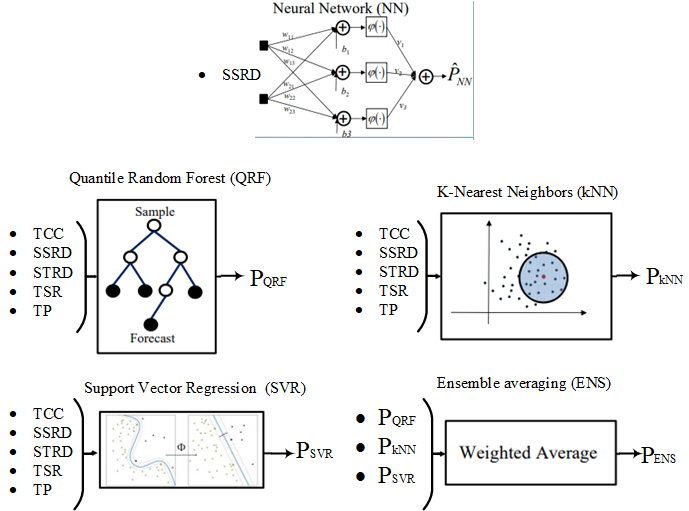}}
\caption{Forecasting methodologies [8].}
\label{THD_analysis}
\centering
\end{figure}

\subsection{Quintile Random Forest (QRF)}
The random forest method is an ensemble of classification models, where each nodes are decision tree. The idea behind using this method is, combining multiple classification methods improves performance than single decision tree. However, challenge is to build uncorrelated trees. For our objective boosting or bagging technique can be used. In this work, we have tried to keep equal no of architecture variables like kNN. Therefore, in this network, we have used 300 decision trees, 5 minimum samples at terminal nodes and MSE method is used for splitting the method. For quantile parameters, we have used 0.4 as per discussed in the literature [8].
\subsection{Support Vector Regression (SVR)}
This method is originally discussed in [8] as SVM, for solving the regression estimation problem. Nowadays, there are many similar existing methods are available [8]. In this work, we have used $v$-SVR method. This method also requires best-tuned meta parameters like aforementioned method for optimal performance. Therefore we have used $v=0.5$, $\gamma=1.25$ for Gaussian kernel function, and regularization parameter $c=1$ as discussed in the literature [8].  
\subsection{Ensemble Averaging (ENS)}
This method ensemble all the aforementioned methods except NN, because of poor performance. The idea behind of this method comes from human intuition. We human beings tend to seek advice before making any important decisions. Similarly, this method combines independent forecasting techniques to obtain another improved forecasting result. Here we have used stacked generalization to combine the above algorithm outputs using the weighted average. Optimal weighted are computed by minimizing the squared error between the each predicted powers and the actual generated powers using pseudo inverse.
\section{Results}

In this section, we have compared predicted power versus actual power. To do this we have used the GEFCom2014 data. This data contains 12 independent meteorological variables cross-ponding to three solar panels, which are situated in different zones. The detailed information of the data is publicly available [9]. For simplicity, we have used Zone1 data from year 2013 for training and validating the models. Then randomly seven days data from Feb 2014 are selected for testing the models. The detailed comparison results are presented in Fig. 2. It can be observed that NN performs very poor during the test, the possible reason might be due to strong dependence of power with other meteorological variables. In this case, we have only taken Irradiance variable as input to predict the output power. Therefore NN performs very poorly than other methods. However, it can be observed that kNN, QRF, and SVR perform well during the testing. To further compare, we have also plotted the nMAE error of 24 hours [8]. From Fig. 3, It is found that at 20, 25 and 26 Feb QRF performed well, whereas SVR performed well on 21 Feb and ENS performance is well on 22, 23 and 24 Feb. The numerical data is also presented in Table I and weekly error is calculated. It is observed that the ENS weekly nMAE is 5.53\%.  
\begin{figure}[h!]
\centerline{\includegraphics[width=0.6\textwidth]{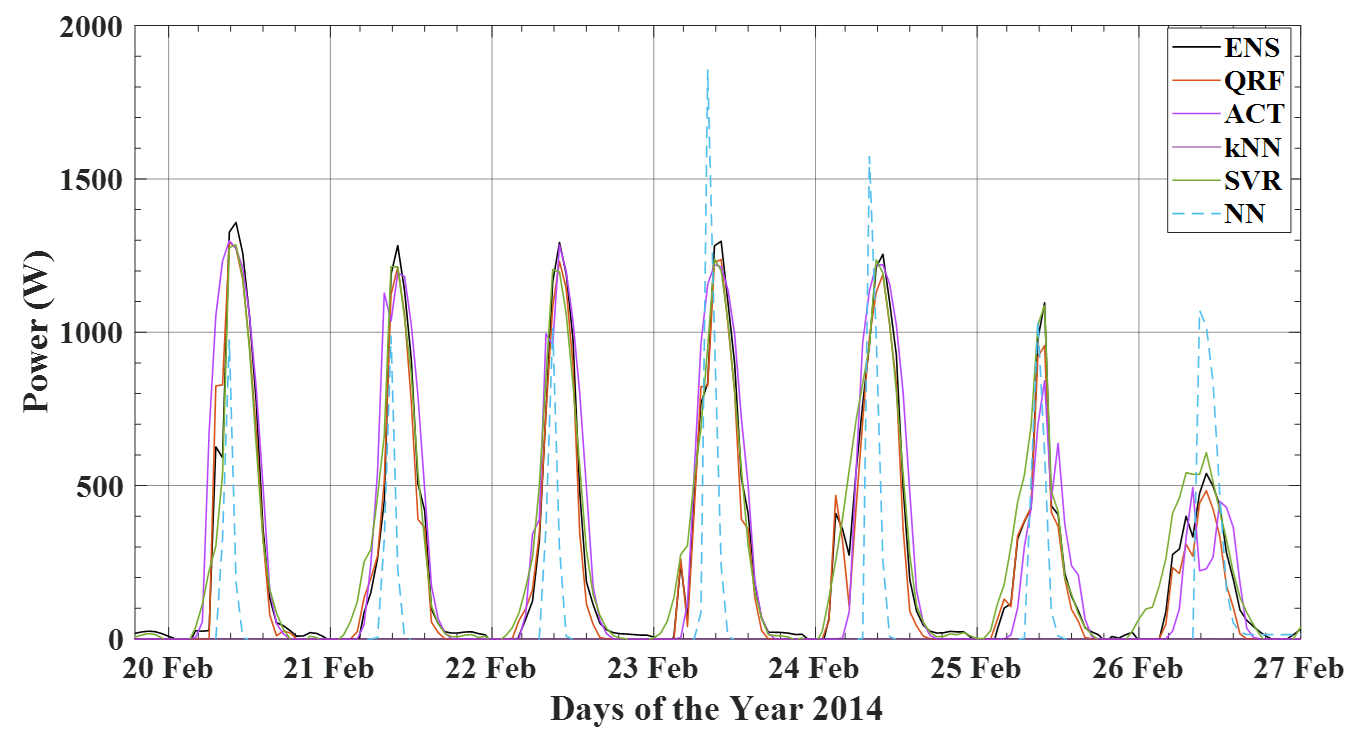}}
\caption{Generated power vs. predicted power in seven following days.}
\label{THD_analysis}
\centering
\end{figure} 

\begin{figure}[h!]
\centerline{\includegraphics[width=0.6\textwidth]{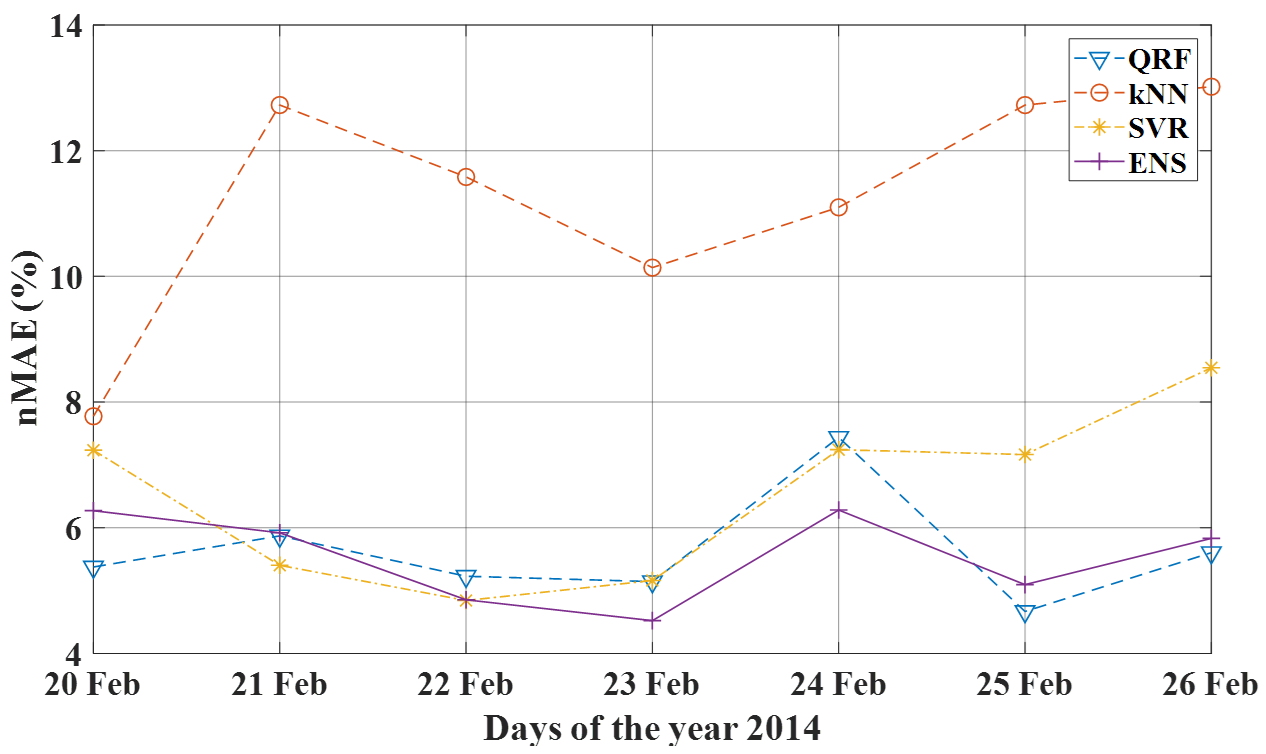}}
\caption{Daily error (nMAE) obtained by each forecasting methodology.}
\label{THD_analysis}
\centering
\end{figure} 

\begin{table}[htb]
\caption{nMAE obtained by each forecasting methodology.}
\begin{center}
\begin{tabular}{c|cccc}
\hline

\hline\hline %inserting double-line 
 \multicolumn{1}{c}{nMAE (\%) }
 &\multicolumn{1}{c}{QRF}
 &\multicolumn{1}{c}{kNN}
 &\multicolumn{1}{c}{SVR}
 & \multicolumn{1}{c}{ENS} \\

\hline 

 & 5.37 & 7.77 & 7.23 & 6.27\\

& 5.87 & 12.72 & 5.40 & 5.92\\
& 5.23 & 11.58 & 4.84 & 4.85\\
& 5.14 & 10.14 & 5.16 & 4.52\\
Daily Error (\%) & 7.44 & 11.09 & 7.24 & 6.28\\
& 4.67 & 12.72 & 7.16 & 5.10\\
& 5.60 & 13.02 & 8.54 & 5.83\\
\hline
Weekly Error (\%)& \textbf{5.61} & \textbf{11.29} & \textbf{6.51} & \textbf{5.53}\\
\hline\hline
\end{tabular}
\label{tab1}
\end{center}
\end{table}

\section{Conclusion}
Despite available of forecasting techniques in literature, performance comparison of NN, QRF, kNN and SVR are missing on publicly available datasets. In this report, we have done comparison analysis of a proposed ENS technique and above-mentioned techniques using publicly available dataset GEFCom2014. It is observed that weekly nMAE of the QRF, kNN, SVR, and ENS during testing is 5.61\%, 11.29\%, 6.51\%, and 5.53\% respectively. Therefore, it can be concluded that the ENS performance is better than state of the art. However, the proposed ENS technique is more complex than individual methodologies due to ensemble of all the mentioned techniques.
\section{Future Work}
Due to limitation of time, several things are not included in this report such as 1) During the training of the QRF, kNN and SVR network, we have only taken 5 variables out of 12 meteorological variables to reduce the complexity of network. The selection of the variable was purely based on intuition, which needs to be justified. 2) During testing of the techniques, we have only taken 1-week data to compare the performance among techniques, which should be done for 1 year. 3) Complexity analysis of the proposed technique needs to be done.

% or


\ifCLASSOPTIONcaptionsoff
  \newpage
\fi


\begin{thebibliography}{00}
\bibitem{b1} R. K. Behera, R. Kumar, S. M. Bellala and P. Raviteja, ``Analysis of electric vehicle stability effectiveness on wheel force with BLDC motor drive," 2018 IEEE International Conference on Industrial Electronics for Sustainable Energy Systems (IESES), 2018, pp. 195-200.
\bibitem{b2} R. Kumar and R. K. Behera, ``A Low Cost Distributed Solar DC Nanogrid: Design and Deployment with Remote Monitoring Unit," 2018 20th National Power Systems Conference (NPSC), 2018, pp. 1-6.

\bibitem{b3} R. Kumar, S. Sneha and R. K. Behera, ``Controller Gain Impact on Islanded dc Microgrid Stability with Constant Power Load," 2018 IEEE International Conference on Power Electronics, Drives and Energy Systems (PEDES), 2018, pp. 1-6.

\bibitem{b4} R. Kumar, M. Rajaput, N. Deshmukh, D. More and S. Anand, ``Partial Unfolding Scheme for Grid Feeding Transformerless PV Inverter," 2018 IEEE International Conference on Power Electronics, Drives and Energy Systems (PEDES), 2018, pp. 1-6.

\bibitem{b5} 
R.A. Verzijlbergh, L.J. De Vries, G.P.J. Dijkema, P.M. Herder, ``Institutional challenges caused by the integration of renewable energy sources in the European electricity sector," Renewable and Sustainable Energy Reviews, Volume 75, 2017, Pages 660-667.

\bibitem{b6}
Alberto Dolara, Sonia Leva, Giampaolo Manzolini, ``Comparison of different physical models for PV power output prediction, Solar Energy," Volume 119, 2015, Pages 83-99.

\bibitem{b7}
Y. Huang, J. Lu, C. Liu, X. Xu, W. Wang and X. Zhou, ``Comparative study of power forecasting methods for PV stations," 2010 International Conference on Power System Technology, Hangzhou, 2010, pp. 1-6.
\bibitem{b8}
L. Gigoni et al., ``Day-Ahead Hourly Forecasting of Power Generation From Photovoltaic Plants," in IEEE Transactions on Sustainable Energy, vol. 9, no. 2, pp. 831-842, April 2018.
\bibitem{b9}
``Global Energy Forecasting Competition 2014 Probabilistic Solar Power Forecasting," Global Energy Forecasting Competition 2014 Probabilistic Solar Power Forecasting. [Online]. Available: https://www.crowdanalytix.com/contests/global-energy-forecasting-competition-2014-probabilistic-solar-power-forecasting. [Accessed: 09-Nov-2019].




\end{thebibliography}
\end{document}